\documentclass[english,aps,preprint,superscript]{revtex4-1}
\usepackage[T1]{fontenc}
\usepackage[utf8]{inputenc}
\usepackage{babel}
\usepackage{amsmath}
\usepackage{amssymb}
\usepackage{graphicx}
\usepackage{csquotes}
\usepackage[unicode=true,pdfusetitle,
 bookmarks=true,bookmarksnumbered=false,bookmarksopen=false,
 breaklinks=false,pdfborder={0 0 1},backref=false,colorlinks=false]
 {hyperref}

\usepackage{placeins}

\makeatletter

\@ifundefined{textcolor}{}
{%
 \definecolor{BLACK}{gray}{0}
 \definecolor{WHITE}{gray}{1}
 \definecolor{RED}{rgb}{1,0,0}
 \definecolor{GREEN}{rgb}{0,1,0}
 \definecolor{BLUE}{rgb}{0,0,1}
 \definecolor{CYAN}{cmyk}{1,0,0,0}
 \definecolor{MAGENTA}{cmyk}{0,1,0,0}
 \definecolor{YELLOW}{cmyk}{0,0,1,0}
}

\@ifundefined{date}{}{\date{}}
\newcommand{\angstrom}{\mbox{\normalfont\AA}}

\makeatother

\begin{document}

\title{Structural and electronic transformation in low-angle twisted bilayer graphene}

\author{Fernando Gargiulo}
\affiliation{Institute of Physics, Ecole Polytechnique Fédérale de Lausanne (EPFL),\\
CH-1015 Lausanne, Switzerland}
\author{Oleg V. Yazyev}
\email{oleg.yazyev@epfl.ch}
\affiliation{Institute of Physics, Ecole Polytechnique Fédérale de Lausanne (EPFL),\\
CH-1015 Lausanne, Switzerland}

\date{\today}

\begin{abstract}


Experiments on bilayer graphene unveiled a fascinating realization of stacking disorder where triangular domains with well-defined Bernal stacking are delimited by a hexagonal network of strain solitons. Here we show by means of numerical simulations that this is a consequence of a structural transformation of the moir\'{e} pattern inherent of twisted bilayer graphene taking place at twist angles $\theta$ below a crossover angle $\theta^{\star}=1.2^{\circ}$. The transformation is governed by the interplay between the interlayer van der Waals interaction and the in-plane strain field, and is revealed by a change in the functional form of the twist energy density. This transformation unveils an electronic regime characteristic of vanishing twist angles in which the charge density converges, though not uniformly, to that of ideal bilayer graphene with Bernal stacking. On the other hand, the stacking domain boundaries form a distinct charge density pattern that provides the STM signature of the hexagonal solitonic network.


\end{abstract}
\maketitle


Bilayer graphene (BLG) shares many of the properties of monolayer graphene while also showing a number of pronounced differences.
For instance,  its equilibrium structural configuration 
reveals the massive nature of its charge carriers  \cite{castro_neto_electronic_2009},
the possibility of inducing a tunable band gap by applying a transverse
electric field  \cite{ohta_controlling_2006,oostinga_gate-induced_2008,zhang_direct_2009}
and quantum Hall valley ferromagnetism  \cite{lee_chemical_2014}.
These properties are a result of the coupling between the two layers.

In order to describe the atomic structure of bilayer graphene the relative position of the two layers has to be defined. In many situations it is sufficient to specify a unique interlayer displacement vector that defines the stacking configuration. As a general property of graphitic structures, the low-energy configuration is represented by the Bernal stacking \cite{bernal_structure_1924,bacon_interlayer_1951}. However, the stacking configuration is not immune to disorder which can manifest, for example, in boundaries that connect two domains with energetically degenerate yet topologically inequivalent stacking configurations, AB and BA  \cite{lalmi_flower-shaped_2014,lin_ac/ab_2013,butz_dislocations_2014,gong_reversible_2013}.
Such stacking domain boundaries are realized by strain solitons, which are segments with a characteristic width
where the strain that arises from interfacing two inequivalent stacking domains is confined. 
Recent studies have shown that strain solitons can be displaced by the action of a scanning tunneling microscope tip, but do not vanish due to their
topological nature \cite{yankowitz_electric_2014,lalmi_flower-shaped_2014}.
From the theoretical point of view, the two-dimensional extension of the Frenkel-Kontorova model predicts the emergence of strain solitons with a typical width of a few nanometers \cite{popov_commensurate-incommensurate_2011} while their density is defined by the twist angle.

In other situations the stacking configuration cannot be uniquely defined on the whole surface
of the sample since the two layers cannot be superimposed by a rigid in-plane
shift.
This is the case of twisted bilayer graphene where one layer is rotated relative to another, a system that has been widely reported in samples grown epitaxially or by chemical vapor deposition \cite{hass_why_2008,li_observation_2010,brown_twinning_2012,lu_twisting_2013,kim_coexisting_2013,beechem_rotational_2014,razado-colambo_nanoarpes_2016}. The two rotated layers form a typical moir\'{e} superlattice that has been imaged by means of transmission electron microscopy (TEM) and scanning tunneling microscopy (STM). 
Importantly, strain solitons in bilayer graphene can form a hexagonal
network that delimits triangular domains with inequivalent AB and BA Bernal stacking \cite{alden_strain_2013,lalmi_flower-shaped_2014,jiang_soliton-dependent_2016}. Remarkably, as pointed out in Ref.~\cite{alden_strain_2013}, such structures are topologically equivalent to twisted BLG.

In this work, we have investigated by means of numerical simulations twisted BLG and show that for twist angles  $\theta$ above a crossover angle  $\theta^{\star}=1.2^{\circ}$ the equilibrium structures do not differ substantially from a rigid twist
of the two layers, while for $\theta$ below $\theta^{\star}$ a crossover into a different regime takes place. In this regime, the equilibrium configuration consists of a triangular lattice of alternating AB and BA stacking domains
separated by shear strain solitons that form a hexagonal network.
The electronic structure is profoundly affected by the emergence of this structural phase and exhibits characteristic
features determined by the local stacking order.
In contrast to the picture valid for low-angle rigidly twisted BLG that predicts low-energy
states localized in AA regions \cite{trambly_de_laissardiere_localization_2010,trambly_de_laissardiere_numerical_2012,uchida_atomic_2014}, we find that the charge density in AB and BA domains resembles that of ideal Bernal-stacked bilayer graphene.
On the other hand, the stacking domain boundaries form a distinct charge density pattern that provides the scanning tunneling microscopy signature of the hexagonal solitonic network.

\section{Results}

\subsection{Structural transformation in low-angle twisted bilayer graphene}

\begin{figure}[!t]
\includegraphics[width=14cm]{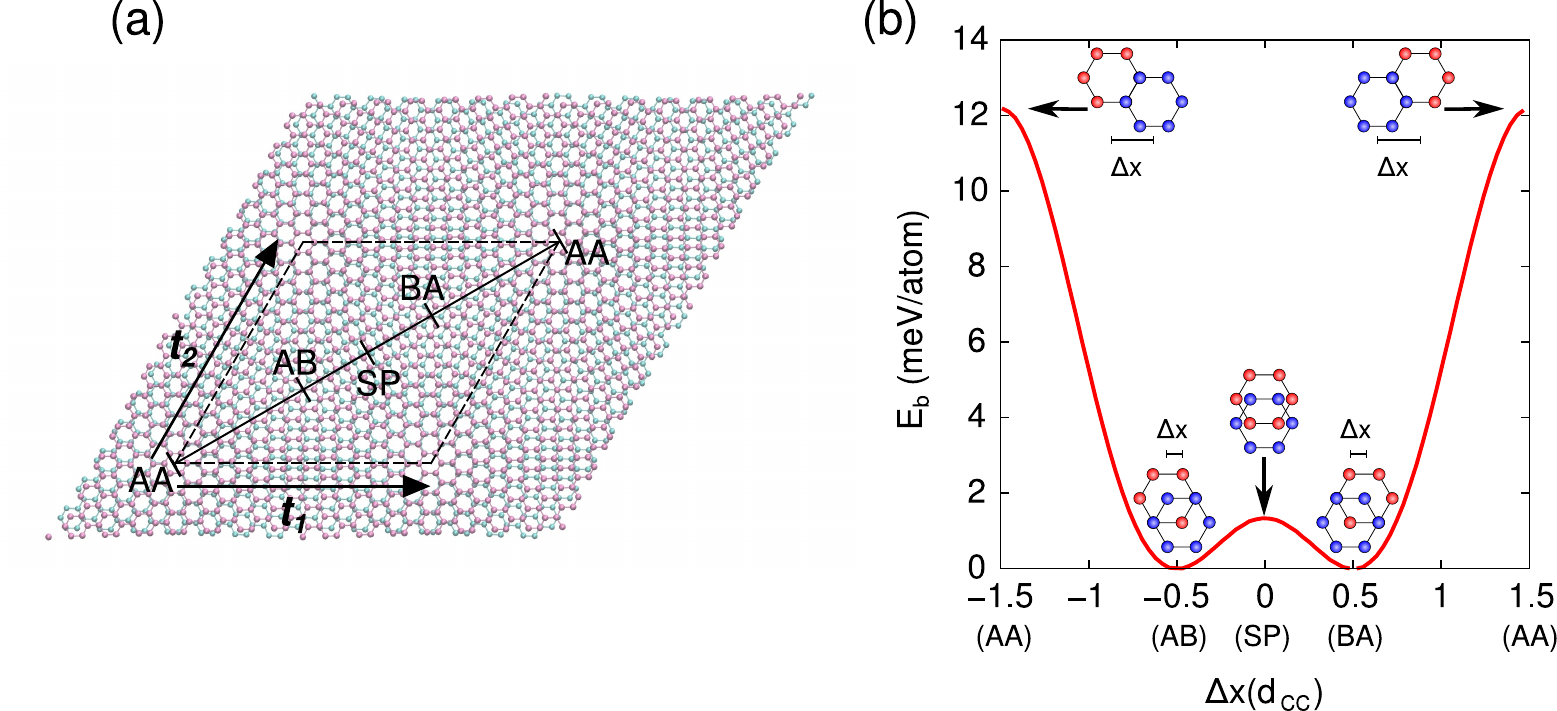}
\protect\caption{\label{fig:Energy landscape}{Interlayer interactions in twisted bilayer
graphene.}  \textbf{a} Ball-and-stick representation of a model of twisted bilayer
graphene characterized by twist angle $\theta=4.4^{\circ}$ and moir\'{e} periodicity $L=3.2~\mathrm{nm}$. The moir\'{e} supercell is highlighted by dashed lines. Along the black
line the stacking order evolves through AA, AB, SP, BA, and AA configurations defined
in  \textbf{b}.  \textbf{b} Interlayer binding energy of bilayer
graphene $E_{b}$ calculated within DFT+vdW as a function of
interlayer lateral displacement $\Delta x$ (see ball-and-stick schemes where
atoms with different colors belong to opposite layers). The interlayer
distance is fixed to $\Delta z=3.412\,\angstrom$. The energy reference $E_{b}\left(\Delta x=\pm0.5~d_{\mathrm{CC}}\right)$ corresponds
to AB/BA stacking configuration.}
\end{figure}

The underlying physical mechanism responsible for the structural transformation occurring in twisted BLG in the limit of small twist angle is
the interplay between van der Waals forces, responsible for the interaction between the
two graphene layers, and in-plane elasticity forces.
As shown in Fig.~\ref{fig:Energy landscape}(a), along the diagonal of a
moir\'{e} supercell of twisted BLG the local stacking
evolves through the high-symmetry configurations AA, AB, SP, BA, and AA (see Fig.~\ref{fig:Energy landscape}(b)
for naming conventions).
The binding energy is minimal for AB/BA stacking configuration,
whereas AA configuration corresponds to the maximum of the potential energy surface, see Fig.~\ref{fig:Energy landscape}(b). The local energy maximum in the middle of the path connecting AB and BA corresponds to an additional high-symmetry stacking configuration commonly referred to as SP.
This potential energy landscape leads to in-plane forces that displace atoms in order to maximize the area of AB/BA stacking domains. On the other hand, the in-plane atomic rearrangement is hindered by the strain caused by the atomic displacement itself. The ultimate equilibrium structures result from the competition between the minimization
of the interlayer energy and the reaction of the strain field \cite{shuyang_twisted_2016}.

In our simulations, we investigate the equilibrium structure of twisted bilayer graphene
by treating atomic interactions using a classical potential. Previous
DFT studies of twisted BLG have been performed within LDA or GGA functionals that disregard dynamical charge correlations responsible
for the van der Waals (vdW) interaction \cite{hass_why_2008,trambly_de_laissardiere_localization_2010,uchida_atomic_2014}. As we aim to treat models with up to $N=3 \times 10^{5}$ atoms,
\textit{ab initio} calculations become prohibitive. Currently
available implementations of classical potentials for carbon do not
reproduce correctly the interlayer energy of layered structures based on $\mathrm{sp^{2}}$-hybridized
carbon atoms    \cite{brenner_second-generation_2002,los_intrinsic_2003,spanu_nature_2009,lebegue_cohesive_2010,reguzzoni_potential_2012}.
In order to describe correctly the interlayer interaction, we define a new potential $V_{\mathrm{LCBOP/KC}}=V_{\mathrm{SR}}+V_{\mathrm{LR}}$
formed by a short range contribution, $V_{\mathrm{SR}}$, inherited
from the LCBOP  potential  \cite{los_intrinsic_2003} and a long-range registry-dependent
contribution, $V_{\mathrm{LR}}$, described by a reparametrized version
of the Kolmogorov-Crespi potential    \cite{kolmogorov_registry-dependent_2005}.
The parameters of $V_{\mathrm{LR}}$ have been fitted in order to reproduce several observables calculated within DFT+vdW (see Section Methods and Supplementary Materials).

Notably, our DFT+vdW calculations reproduce well the equilibrium interlayer distance of graphite $3.36~\angstrom$ and the in-plane bond length $1.42\,\angstrom$. For bilayer graphene, we have found the atomic bond-length $d_{\mathrm{CC}}=1.419\,\angstrom$ and the equilibrium interlayer
distances of AB, AA, and SP stacking configurations, respectively, $\Delta z_{\mathrm{AB}}=3.412\,\mathrm{\angstrom}$,
$\Delta z_{\mathrm{AA}}=3.599\,\mathrm{\angstrom}$, and $\Delta z_{\mathrm{SP}}=3.439\,\mathrm{\angstrom}$. We have found that the interlayer energies for AA and SP stacking configurations calculated at $\Delta z=3.412~\angstrom$ are, respectively, $E_b^{\mathrm{AA}}=12.2~\mathrm{meV/atom}$ and $E_b^{\mathrm{SP}}=1.33~\mathrm{meV/atom}$ relative to AB configuration. 

Supercells of twisted bilayer graphene have been built according
to the rules derived by imposing commensurability conditions \cite{lopesdossantos_graphene_2007,shallcross_quantum_2008}. In particular,
one class of supercells is defined by an integer $w$ that determines
the supercell periodicity vectors $\mathbf{t}_{1}=w\mathbf{a}_{1}+(w+1)\mathbf{a}_{2}$
and $\mathbf{t}_{2}=-(w+1)\mathbf{a}_{1}+(2w+1)\mathbf{a}_{2}$, with $\mathbf{a}_{1}$
and $\mathbf{a}_{2}$ ($\left|\mathbf{a}_{1/2}\right|=\sqrt{3}d_{\mathrm{CC}}$)
being the crystal vectors of the graphene honeycomb lattice, and the
corresponding twist angle is defined by $\cos\theta=(3w^{2}+3w+1/2)/(3w^{2}+3w+1)$   \cite{lopesdossantos_graphene_2007}.
Vectors $\mathbf{t}_{1}$ and $\mathbf{t}_{2}$ form a $60^{\circ}$ angle
and the moir\'{e} pattern has $\mathrm{C}_{3}$ symmetry, see Fig.~\ref{fig:Energy landscape}(a).
Notably, in the limit $w\rightarrow\infty$ the twist angle and the
supercell linear size are inversely proportional: $\theta^{-1} \propto \left|\mathrm{\mathbf{t}}_{1}\right|=L$.
We have performed atomic structure relaxation of models with index $w$ up to $160$ corresponding to $\theta=0.206^{\circ}$,
$L=68.4\,\mathrm{nm}$, and total number of atoms $N=309124$.

\begin{figure}
\centering{}\includegraphics[width=15cm]{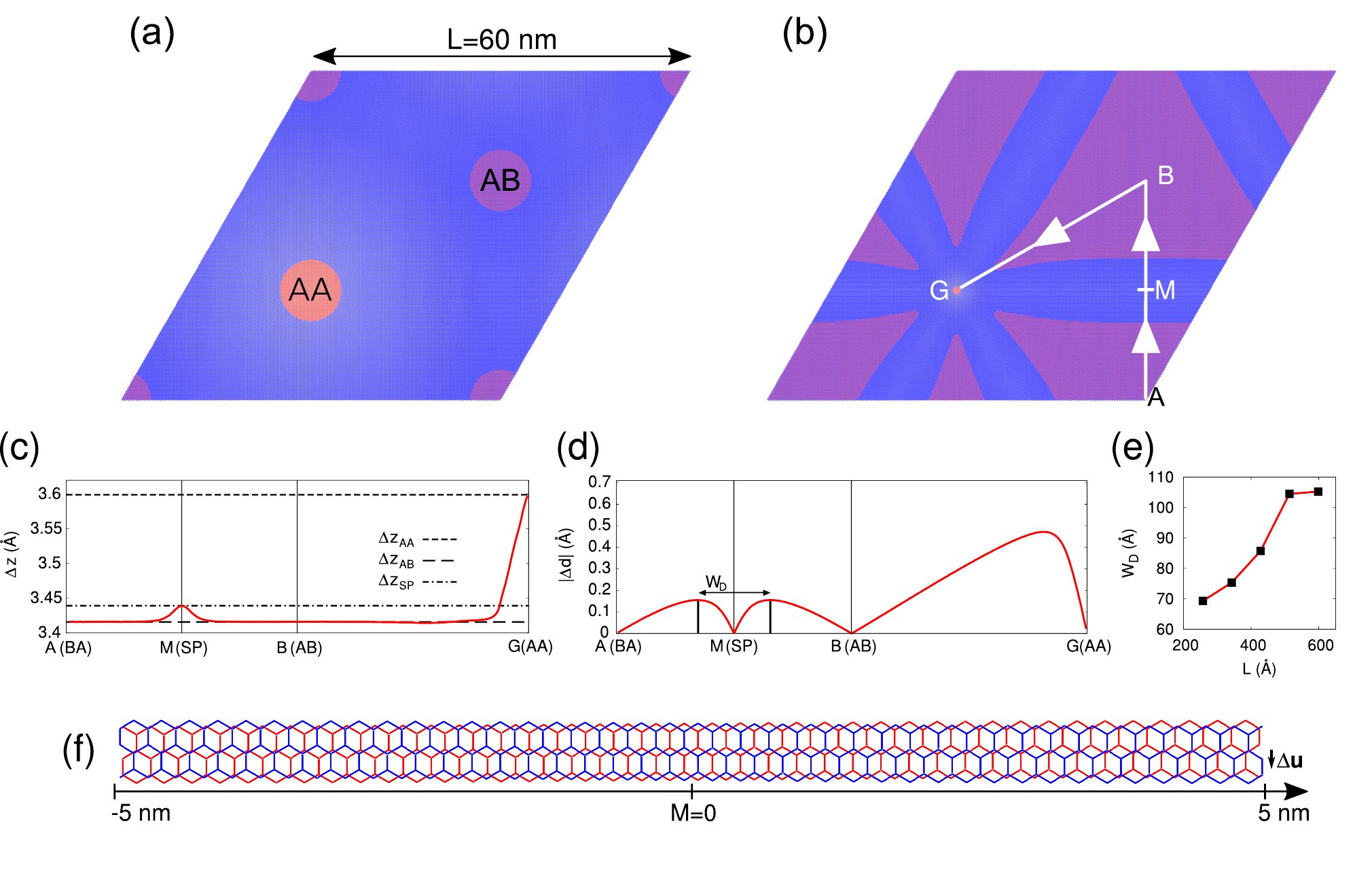}
\protect\caption{\label{fig:Relaxation}{Atomic relaxation of a model of twisted bilayer graphene
characterized by $\theta=0.235^{\circ}$ and $L=\left|\mathbf{t}_{1}\right|=59.8\,\mathrm{nm}$
($w=140$).}  \textbf{a} Representation of the initial model with rigidly rotated
layers. Pairs of atoms in opposite layers whose lateral positions
are closer than $0.2\,\mathrm{\angstrom}$ are colored in
red, the remaining in blue. The AA stacking regions (within $0.2\,\mathrm{\angstrom}$ tolerance) contain only red atoms,
whereas atoms in AB stacking regions are alternatively colored in blue and red, and hence such regions appear in purple.
Regions with neither AA nor AB stacking are blue.
This representation allows distinct stacking domains to be recognized
at a glance.  \textbf{b} Representation of the relaxed structure with the
same color-coding procedure as in  \textbf{a}. \textbf{c} Interlayer distance $\Delta z$
for the relaxed system along the path $\mathrm{AMBG}$ defined
in  \textbf{b}.
\textbf{d} Absolute magnitude of the atomic displacement driven by the in-plane strain along the path $\mathrm{AMBG}$.
\textbf{e} Dependence of $W_{\mathrm{D}}$, defined as the distance between the two symmetric maxima with respect to $\mathrm{M}$, on the moir\'{e} periodicity $L$.  \textbf{f} Shear soliton separating AB
and BA domains arising from structural relaxation, with $\Delta \mathbf{u}$ representing the shear strain vector.}
\end{figure}

Figs.~\ref{fig:Relaxation}(a,b) show a model of rigidly twisted BLG with $\theta=0.235^{\circ}$ and the corresponding equilibrium structure resulting from relaxation.
Upon relaxation, AB(BA) regions have extended and transformed in approximately triangular domains with $\simeq40\,\mathrm{nm}$ side, while AA regions reduce to much smaller, nanometer-scale patches in order to minimize the energy penalty payed with respect to Bernal stacking, see Fig.~\ref{fig:Energy landscape}(b).
The segments separating AB/BA domains form a hexagonal lattice with
vertices corresponding to the AA regions. The modulus of the atomic in-plane displacement upon relaxation,
$\Delta d$, along a high-symmetry path is shown in Fig.~\ref{fig:Relaxation}(d).
Along the line connecting point $A$ or $B$ with point $\mathrm{M}$, carbon atoms
increasingly displace to restore AB/BA stacking.
Because of the opposite value of $\Delta d$ along segments $\rm AM$ and $\rm BM$, strain
concentrates in the vicinity of $\mathrm{M}$. When the energy gained
by restoring AB or BA stacking order is compensated by the local strain energy,
$\Delta d$ reaches a maximum before abruptly vanishing. The distance $W_{\mathrm{D}}$ between the two symmetric
maxima converges to a constant value $W^{\star}_\mathrm{D}=10.5\,\mathrm{nm}$ upon increasing the moir\'{e} periodicity $L$. This illustrates
the emergence of a shear strain soliton separating two stacking
domains. As shown in Fig.~\ref{fig:Relaxation}(f), the modulus of shear strain vector, $\Delta\mathbf{u}$, corresponds to one carbon-carbon bond length, $\left|\Delta\mathbf{u}\right|=1.42\,\angstrom$.  A similar
reasoning is valid for the path $\mathrm{BG}$. Additionally,
atoms displace in the out-of-plane direction as shown in Fig.~\ref{fig:Relaxation}(c).
Out-of-plane relaxation of twisted BLG has been investigated within
DFT/LDA for $2^{\circ} < \theta$ finding out-of-plane corrugation at low twist angles \cite{uchida_atomic_2014}. A work based on a classical potential investigated twisted BLG in the regime $0.46^{\circ}<\theta<2.1^{\circ}$  \cite{wijk_relaxation_2015}.
Similarly to what was reported in Ref.~\cite{wijk_relaxation_2015}, we find that $\Delta z$ approaches the value $\Delta z_{\mathrm{AB}}$ in AB/BA stacking domains giving rise to the plateaus seen in Fig.~\ref{fig:Relaxation}(c). Moreover, $\Delta z$ adapts to $\Delta z_{\mathrm{SP}}$ and $\Delta z_{\mathrm{AA}}$,
respectively, at $\mathrm{M}$ and $\mathrm{G}$, consistent
with the local stacking configurations \cite{jain_structure_2017}. Therefore, this leads to a small corrugation (tilt angle of the normal vectors $\alpha<0.2^{\circ}$) at the location of shear solitons as well as their junctions \cite{shuyang_twisted_2016, jain_structure_2017}. 

\begin{figure}[!th]
\includegraphics[width=14cm]{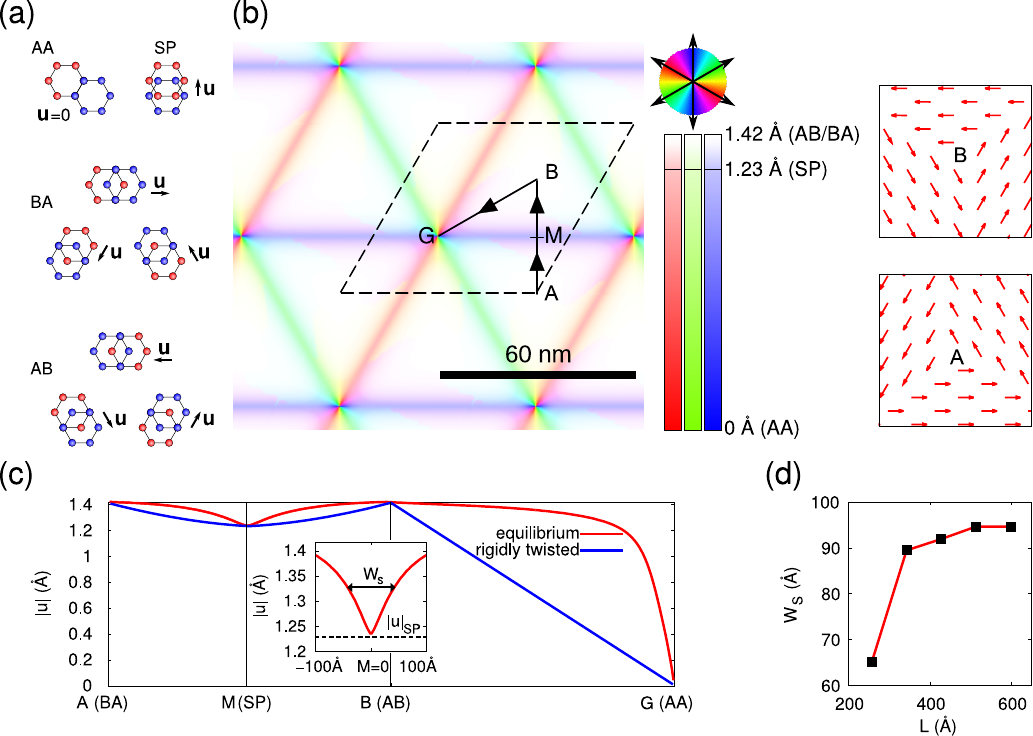}\protect\caption{\label{fig:Stacking}{Representation of the stacking vector field $\mathbf{u}$
of a model of twisted BLG}. \textbf{a} Illustration of the stacking vector field $\mathbf{u}$ for high-symmetry configurations.
\textbf{b} Color-coded representation of field $\mathbf{u}$ for the equilibrium configuration of a twisted BLG model with $\theta=0.235^{\circ}$, $L=59.8\,\mathrm{nm}$ ($w=140$). Hue and saturation at each point represent, respectively, the direction and
the intensity of the local value of $\mathbf{u}$. Fully saturated
colors correspond to AA stacking configuration ($\mathbf{\left|u\right|=}0~\mathrm{\angstrom}$),
white regions (vanishing saturation) correspond to AB stacking configuration ($\mathbf{\left|u\right|=}1.42~\mathrm{\angstrom}$). SP-stacked soliton centers ($\mathbf{\left|u\right|=}1.23\mathbf{\,\mathrm{\angstrom}}$)
are half-saturated. Hue varies with a period of $180^{\circ}$ as
shown in the wind-rose diagram. Lateral panels show the stacking vector field
in the vicinity of B and A.  \textbf{c} Absolute magnitude of $\mathbf{u}$ along the
path $\mathrm{AMBG}$ for equilibrium and rigidly twisted BLG. The full width at half maximum of the dip at $\mathrm{M}$ is referred
to as a $W_{\mathrm{S}}$ and used as definition of the soliton width.
The dependence of $W_{\mathrm{S}}$ on the periodicity $L$ is shown in \textbf{d}.}
\end{figure}

To obtain a deeper insight into the equilibrium configurations of twisted BLG, we have studied the stacking vector field $\mathbf{u}$,
defined as the in-plane component of the minimal shift that has to be
applied to one layer in order to make it coincide locally with the opposite layer,
see Fig.~\ref{fig:Stacking}(a).

Fig.~\ref{fig:Stacking}(b) shows the presence of triangular domains with almost constant stacking $\left|\mathbf{u}\right|=1.42\,\mathrm{\angstrom}$ (white regions). By inspecting the local stacking field around $\mathrm{A}$ and $\mathrm{B}$
(side panels in Fig.~\ref{fig:Stacking}(b)) one can see the confluence
of three orientations of $\mathbf{u}$ differing by $120^{\circ}$. This discontinuity
is trivial as the vector $\mathbf{u}$ for AB(BA) stacking has three
degenerate representations forming $120^{\circ}$ angles with
each other, see Fig.~\ref{fig:Stacking}(a). However, when following a path connecting one stacking domain with another across a shear soliton, (e.g. from point $\mathrm{A}$ to point $\mathrm{B}$), $\mathbf{u}$
rotates by $60^{\circ}$, that is, the stacking configuration changes from AB to
BA or vice versa. The variation $\Delta\mathbf{u}$
($\left|\Delta\mathbf{u}\right|=1.42\,\angstrom$) is parallel to the
strain soliton and coincides with its shear vector. These stacking
domain boundaries are topological defects and $\Delta\mathbf{u}$
is assigned as their topological invariant. In the following, the denominations \enquote{shear
soliton} and \enquote{stacking domain boundary} will be used interchangeably.

As shown in Fig.~\ref{fig:Relaxation}(c), along the path $\mathrm{AMB}$
$\mathbf{\left|u\right|}$ has a minimum at $\mathrm{M}$, corresponding
to SP stacking ($\mathbf{\left|u\right|}=1.23\,\angstrom$). Upon increasing
$L$, the full width at half maximum of this dip, $W_{\mathrm{S}}$, saturates to $W_{\mathrm{S}}^{\star}=9.5\,\mathrm{nm}$, a value close to $W_{\mathrm{D}}^{\star}=10.5\,\mathrm{nm}$. We choose to use $W_{\mathrm{S}}^{\star}$ to define the width of the stacking
domain boundaries since its determination does not require reference to the corresponding rigidly twisted structure. The calculated widths of the solitons are in good accordance with the experiments    \cite{alden_strain_2013,lin_ac/ab_2013,butz_dislocations_2014}.
The vertices of the hexagonal network ($\mathrm{G}$) where six
stacking domain boundaries merge are topological point defects with
$\mathbf{u}=0$ and a non-zero winding number, that is,
$\mathbf{u}$ rotates by $360^{\circ}$ along a closed path encompassing
$\mathrm{G}$    \cite{mermin_topological_1979}. Noteworthy, that in the vicinity of these vertices, due to energetically unfavorable stacking of the latter close to AA, the width of the solitons is smaller than $W_{\mathrm{S}}^{\star}$.

\begin{figure}[!t]
\includegraphics[width=15cm]{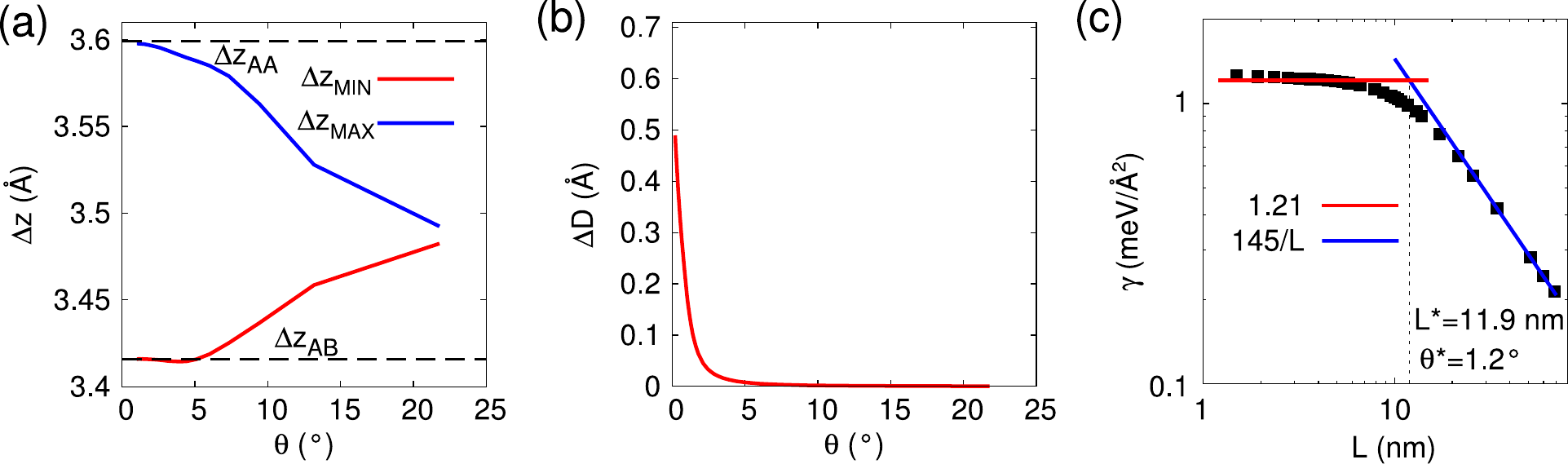}\protect\caption{\label{fig:Transition}{Structural evolution of twisted bilayer graphene in the limit of small twist angle.}  \textbf{a} Largest (smallest) interlayer distance
$\Delta z_{\mathrm{MAX}}(\Delta z_{\mathrm{MIN}})$ as a function of moir\'{e} periodicity $L$.  \textbf{b} Maximum atomic displacement $\Delta D$
as a function of twist angle $\theta$.  \textbf{d} Density of twist energy  $\gamma$
as a function of moir\'e periodicity $L$. Regions with constant and inversely proportional dependence have been fitted, respectively, by red and blue lines, intersecting at $L^{*}=11.9\,\mathrm{nm}$ (corresponding to crossover angle
$\theta^{*}=1.2^{\circ}$).}
\end{figure}

Intuitively, we expect that a transformation involving the creation of strain solitons
takes place when the twisted BLG supercell is larger than $W_{\mathrm{S}}^{\star}$
such that the strain field can be efficiently accommodated.  The dependence
of several observables on $\theta$ (or, equivalently, $L$) reveals further details
of the evolution. As shown in Fig.~\ref{fig:Transition}(a), for
$\theta=21.8^{\circ}$ the distribution of the interlayer distance, $\Delta z$, has a minimal
spread $\Delta z_{\mathrm{MAX}}-\Delta z_{\mathrm{MIN}}\simeq0.01\,\angstrom$. Upon reduction of $\theta$, $\Delta z_{\mathrm{MAX}}$ and $\Delta z_{\mathrm{MIN}}$
increasingly differ and at $\theta\simeq2^{\circ}$ saturate
to $\Delta z_{\mathrm{MAX}}$=$\Delta z_{\mathrm{AA}}$ and $\Delta z_{\mathrm{MAX}}$=$\Delta z_{\mathrm{AB}}$,
consistent with the data shown in Fig.~\ref{fig:Relaxation}(c).
Out-of-plane relaxation competes with the bending rigidity of graphene, estimated as $B_{\mathrm{M}}=1.44\,\mathrm{eV}$    \cite{wei_bending_2013}.
For lower values of $\theta$ in-plane atomic displacements become non-negligible.
Note that the maximum in-plane displacement of individual atoms, $\Delta D$, is bounded from above by a half bond length
$d_{\mathrm{CC}}/2=0.71\,\mathrm{\angstrom}$ since two inequivalent
stacking vectors are connected by $\left|\Delta\mathbf{u}\right|<d_{\mathrm{CC}}$
and the displacement is equally distributed over the atoms in the two layers. The
crossover is underpinned by the change in functional dependence of the density of twist
energy on the moir\'{e} periodicity: $\gamma\left(L\right)=\left(E\left(L\right)-E_{\mathrm{AB}}\right)/A_{\mathrm{S}}$, where $E\left(L\right)$ is the total energy for a supercell
of periodicity $L$, and $E_{\mathrm{AB}}$ is the energy of a AB bilayer graphene
supercell having the same surface area $A_{\mathrm{S}}=L^{2}\sqrt{3}/2$,
see Fig.~\ref{fig:Transition}(c). For small values of $L$, atomic in-plane displacements due to relaxation are negligible and the energy required to introduce a twist arises from those regions whose stacking configuration is not AB/BA, which represent a constant fraction of the supercell surface. Thus, the difference
$E\left(L\right)-E_{\mathrm{AB}}$ is proportional to the surface of the system and the twist energy density equals
a constant: $\gamma\left(L\right)=\gamma_{\mathrm{A}}=1.2\,\mathrm{meV/\angstrom^{2}}$. This has been confirmed by DFT calculations, see Fig.~S1(c) of Supplementary Informations.
On the other hand, for large supercells most of whose area is composed of AB and BA stacking domains,
only the soliton network contributes to $E\left(L\right)-E_{\mathrm{AB}}$.
As the width of the solitons asymptotically approaches the constant
value $W_{\mathrm{S}}^{\star}$, the twist energy density is given by $\gamma=3\gamma_{\mathrm{S}}L/A_{\mathrm{S}}\propto1/L$,
where $\gamma_{\mathrm{S}}$ is the energy per soliton unit length
and the factor $3$ counts the number of solitons in the moir\'{e} supercell.
We estimate $\gamma_{\mathrm{S}}=42\,\mathrm{meV}/\mathrm{\angstrom}$.
The crossover length $L^{\star}=\gamma_{\mathrm{S}}/\gamma_{A}=11.9\,\mathrm{nm}$, corresponding to the crossover angle $\theta^{\star}=1.2^{\circ}$, is defined as the
intersection of the constant line and the curve $\propto1/L$ fitting
the two distinct regimes, as shown in Fig.~\ref{fig:Transition}(c).
Finally, we can rigorously answer why the transformation takes place at large moir\'e periodicities. Regardless the values of $\gamma_{\mathrm{S}}$
and $\gamma_{\mathrm{A}}$, the constant \enquote{rigid} regime is favorable
for $L<\gamma_{\mathrm{S}}/\gamma_{A}$, whereas the $\propto1/L$ \enquote{solitonic} regime is favorable
for $L>\gamma_{\mathrm{S}}/\gamma_{A}$. 

\subsection{Electronic structure of twisted bilayer graphene}

The low-energy states of twisted BLG with large
to intermediate twist angles $3^{\circ}\lesssim\theta\lesssim15^{\circ}$
can be described by a model that introduces the coupling
between graphene layers perturbatively   \cite{lopesdossantos_graphene_2007,bistritzer_moire_2011,koshino_electronic_2013}.
This model predicts the existence of low-energy massless Dirac fermions with $\theta$-dependent
Fermi velocity and a pair of Van Hove singularities slightly asymmetric
with respect to the Dirac point. These predictions have been confirmed experimentally \cite{li_observation_2010,luican_single-layer_2011,jung_origin_2015,wong_local_2015}. For
smaller twist angles $1^{\circ}<\theta<3{}^{\circ}$, twisted BLG has been
predicted to develop a flat band responsible for a zero-energy peak
in the density of states (DOS) \cite{trambly_de_laissardiere_localization_2010,uchida_atomic_2014}.
This peak is due to states localized in AA regions as a result of
the super-periodic potential induced by the moir\'{e} pattern. However,
these results cannot be extrapolated to lower values of $\theta$, as we
expect that the structural relaxation suppressing AA-stacked regions strongly affects the
electronic structure. 

We investigate the low-energy electronic properties of the equilibrium
structure of twisted bilayer graphene in the limit $\theta\rightarrow0^{\circ}$
by means of a tight-binding model taking into account $2p_{z}$
orbitals with hopping parameters depending on the distance between orbital centers
as well as the relative orientation of the orbitals. The latter is achieved
by means of the Slater-Koster theory  \cite{slater_simplified_1954}, see \textit{Supplementary Information} for details. 

\begin{figure}
\includegraphics[width=16.5cm]{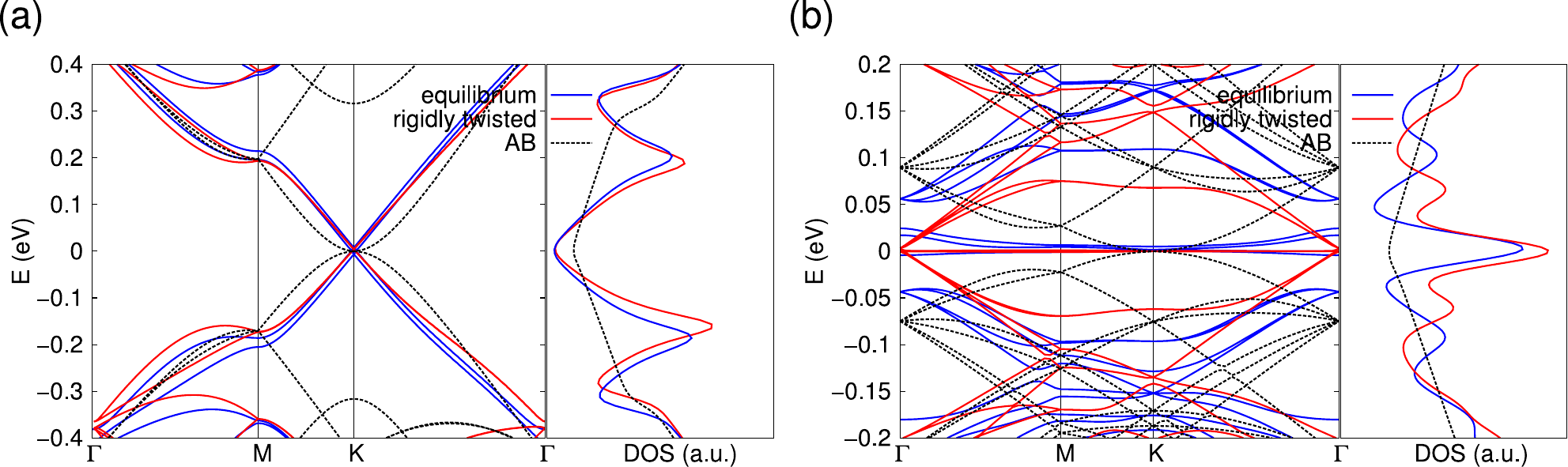}\protect\caption{\label{fig:bands}{Electronic structure of equilibrium and rigidly twisted BLG.}  \textbf{a,b}  Energy bands and density of states (DOS) for two models of twisted BLG characterized by twist angles (\textbf{a}) $\theta=3.8^{\circ}$ ($L=3.6\,\mathrm{nm}$,
$w=8$) and (\textbf{b}) $\theta=1.2^{\circ}$ ($L=12.1\,\mathrm{nm}$,
$w=28$). The energy bands and the DOS calculated for an equivalent supercell of AB graphene are plotted
for comparison. Energies are referenced to the Fermi level.}
\end{figure}

In Fig.~\ref{fig:bands}(a), the band structure for a model with twist angle $\theta=3.8^{\circ}$
shows two degenerate Dirac cones projected onto the $\mathrm{K}$ point
of the supercell Brillouin zone, in contrast to the parabolic dispersion of
AB graphene \cite{lopesdossantos_graphene_2007,shallcross_quantum_2008,hass_why_2008}. A finite coupling between the states in the two Dirac cones is responsible for low-dispersion bands around the M point, whence the appearance of two low-energy Van Hove singularities in the DOS   \cite{lopesdossantos_graphene_2007}. We find that the relaxation has negligible effects on the Dirac fermions,
except for lifting the degeneracy of the low-energy bands. 
As the twist angle decreases, the positions of each Van Hove singularity approaches the Dirac energy, eventually merging at $\theta \simeq 2^{\circ}$. In this regime, the Fermi velocity is zero and the low-energy states are localized in the AA regions.
As shown in Fig.~\ref{fig:bands}(b),
at the crossover angle $\theta^{\star}=1.2^{\circ}$ the DOS for
rigidly twisted BLG shows a triplet of peaks at energies $E\mathcal{\simeq}-0.06$, $0$, and $0.07\,\mathrm{eV}$
that correspond to the flat low-energy bands observable in the band structure. We have found that the
relaxation is responsible for lifting the degeneracy of the central
band and shifting the side peaks further away from the Fermi level, however, without
introducing qualitative modifications of the electronic structure.

\begin{figure}
\includegraphics[width=17cm]{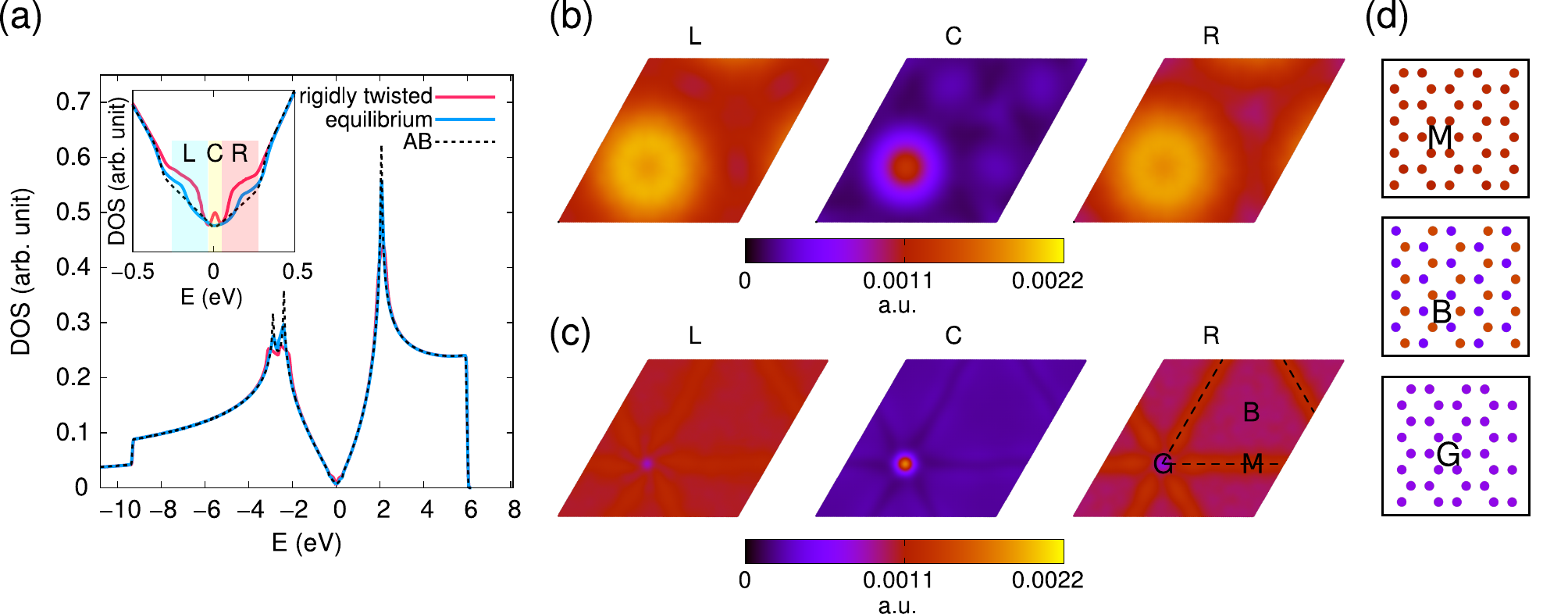}\protect\caption{\label{fig:DOS}{Density of states for
equilibrium and rigidly twisted BLG.}  \textbf{a} Density of states (DOS) as a function of energy $E$ for twisted bilayer graphene
with $\theta=0.235^{\circ}$ ($L=59.8\,\mathrm{nm}$). DOS for equilibrium and rigidly twisted
bilayer graphene are compared. As a reference, the DOS of AB bilayer
graphene ($\theta=0^{\circ}$) is also shown (dashed line). The inset
presents the same data in the energy range $\left[-0.5,0.5\right]\,\mathrm{eV}$.
Regions L, C and R individuate, respectively, the energy intervals
$\left[-0.25,-0.038\right]\,\mathrm{eV}$, $\left[-0.038,0.043\right]\,\mathrm{eV}$,
and $\left[0.043,0.26\right]\,\mathrm{eV}$. Energies are referenced to the Fermi level.  \textbf{b,c} Local density of states (LDOS) integrated
in the energy intervals L, C, and R for  (\textbf{b}) rigidly twisted BLG and for (\textbf{c}) the corresponding equilibrium configuration. Solitons are highlighted by dashed lines. \textbf{d} Same as right panel in  \textbf{c} restricted to 1~$\mathrm{nm^{2}}$ squares centered around
$\mathrm{M}$, $\mathrm{B}$, and $\mathrm{G}$.}
\end{figure}

Fig. \ref{fig:DOS}(a) shows the density of states plot for a model characterized by $\theta=0.235^{\circ}$,
that is, well below the crossover angle $\theta^\star$. The DOS of rigidly twisted BLG
still exhibits a zero-energy peak with two satellite shoulders.
The nature of the low-energy states is revealed by the inspection of local density of states (LDOS) integrated
in the energy regions around each of the three peaks, see Fig. \ref{fig:DOS}(b).
In all cases the charge density is localized in AA regions and
extends on a fraction of surface $\eta\simeq5\%$ for the central
peak and $\eta\simeq25\%$ for satellites, that we found to be largely independent of the moir\'{e}
periodicity $L$.

The picture changes drastically upon in-plane relaxation responsible for the discussed structural transformation. 
The DOS of relaxed twisted BLG is overall closer to AB bilayer graphene and, in particular, the zero-energy peak due to the localized states in the AA region is strongly suppressed (Fig.~\ref{fig:DOS}(a), inset). We note that the localized states are still present and confined to nanometer-size AA regions, as can be seen in the corresponding LDOS maps (compare Figs.~\ref{fig:DOS}(b) and \ref{fig:DOS}(c), central map). The suppression can be explained as follows. 
In the solitonic regime, for twist angles $\theta < \theta^{\star}$, relaxation leads to AA regions of constant area. Therefore, upon decreasing twist angle $\theta$, or equivalently, increasing moir\'e periodicity $L$, the weight of these states in the total DOS decreases as $L^{-2}$. This overall decrease of the zero-energy peak weight is expected to be accompanied by its narrowing as suggested previously \cite{san_jose_helical_2013}. However, our numerical calculations performed on large-scale models do not allow to address adequately this effect. 
The effect of structural relaxation is even more dramatic in the energy ranges corresponding to the side peaks (Fig.~\ref{fig:DOS}(c), maps
L and R).
Indeed, the charge density is partially depleted in AA regions and AB/BA domains show an overall
homogeneous distribution, whereas solitons exhibit a slightly larger charge density. Upon closer investigation of the center of an AB domain (point $\mathrm{B}$ in Fig.~\ref{fig:DOS}(d)) one can observe an alternation
of atoms with high and low charge density. This is typical of graphene layers with AB stacking configuration as demonstrated
by STM images of highly ordered pyrolytic graphite \cite{pong_review_2005}. The reason is the nonequivalence of the two sublattices of AB stacked graphene that reflects different out-of-plane matrix elements for atoms in complementary sublattices.
Charge densities in the soliton regions and at the vertices of the network do not show local variations on the atomic scale (Fig.~\ref{fig:DOS}(d), points $\mathrm{M}$ and $\mathrm{G}$). This is consistent with the fact that the stacking configurations SP and AA found in the solitons and in the vertices, respectively, preserve the sublattice equivalence.

\section{Conclusions}

We investigated the equilibrium low-energy structure of twisted
bilayer graphene in the limit of vanishing twist angle (down to $\theta \approx 0.2^{\circ}$)
by means of simulations based on a classical potential, which is capable of
describing the dependence of the interlayer binding energy on the relative position of the two layers. Carbon atoms displace in order to maximize
the area of energetically favorable AB/BA stacking domains that
assume a triangular shape. The in-plane strain field, thus, appear confined in a
hexagonal network of shear solitons of width $W_{\mathrm{S}}\simeq9.5\,\mathrm{nm}$, that delimit alternating AB and BA stacking domains.
This structural transformation is continuous and takes place at twist angles below the crossover value $\theta^{\star}=1.2^{\circ}$, at which the moir\'{e} superlattice period exceeds the soliton width $W_{\mathrm S}$. 
In the limit $\theta\rightarrow0^{\circ}$, the equilibrium
structure of twisted BLG converges to that of ideal AB-stacked BLG ($\theta=0^{\circ}$). However, the convergence is not uniform
in the sense that the relative abundance of the AB-stacking regions approaches
$1$, but the soliton network due to its topological
nature vanishes only at $\theta=0^{\circ}$. 
On the other hand, twisted BLG as such is not stable with respect
to AB-stacked BLG ($\theta=0^{\circ}$) and its existence is governed by kinetic bottlenecks.

This fact has major consequences on the low-energy electronic states
of the moir\'{e} superlattice. Differently from the
range $1.2^{\circ}<\theta<2^{\circ}$ where the DOS of twisted BLG hosts three
low-energy peaks due to flat bands of states localized on AA regions,
equilibrium structures of twisted BLG with $\theta<1.2^{\circ}$ show a DOS resembling that of AB bilayer graphene with a low-energy charge density distribution that can be directly inferred from the local stacking. The charge density is uniform overall in AB/BA-stacked domains, but shows a strong imbalance between the two inequivalent sublattices in each layer. Conversely, the solitons and the network vertices show no breaking of sublattice symmetry.
This distinctive pattern enables the identification of the stacking domain boundaries by means of STM experiments.
Analogously to the stacking, the relative extent of the regions where the charge distribution differs from that of AB bilayer graphene asymptotically vanishes for $\theta\rightarrow0^{\circ}$.

\section{Methods}

\noindent
\textbf{DFT calculations.} We have employed the rVV10 functional that treats exchange-correlation energy within the GGA and includes a non-local van der
Waals (vdW) contribution \cite{vydrov_nonlocal_2010,sabatini_nonlocal_2013} implemented in QUANTUM ESPRESSO \cite{giannozzi_quantum_2009}. The ion-electron
interaction has been described by means of ultra-soft pseudopotentials \cite{murray_investigation_2009}. Energy
cutoff for wavefunctions and charge density have been set, respectively,
to $E_{\mathrm{wf}}=80\,\mathrm{Ry}$ and $E_{\mathrm{rho}}=574\,\mathrm{Ry}$ and the
Brillouin zone has been sampled with a $16\times16\times1$ Monkhorst-Pack
kpoint grid. All computational parameters and technical details are
listed and discussed in \textit{Supplementary Information}. \\
\textbf{Classical potential simulations.} We have employed the long-range
carbon bond order potential (LCBOP) replacing the original long-range
contribution by a reparametrized version of Kolmogorov-Crespi registry-dependent
potential \cite{kolmogorov_registry-dependent_2005} fitted to match
the DFT/rVV10 values of the interlayer binding energy as a function of
interlayer distance and relative shift. The fit has been performed with
the non-linear minimizer provided by DAKOTA code \cite{adams_dakota_2009}.
Additional details about the fit procedure and the resulting parameters can be found in\textit{ Supplementary
Information.} The optimized potential has been implemented in LAMMPS to perform
energy minimizations \cite{plimpton_fast_1995,lammps_2015}.\\
\textbf{Electronic structure calculations.} We have considered a Slater-Koster \cite{slater_simplified_1954} tight-binding model taking into account $2p_{z}$
orbitals for carbon atoms with hopping parameters depending on the distance between orbital centers
as well as the relative orientation of the orbitals.
This is particularly important in order to describe correctly the interactions
in the soliton region where the relative position of carbon atoms in opposite
layers changes continuously. Since the equilibrium structures show only weak
corrugation, in-plane orbital interactions are predominantly
of the $pp\pi$ type. For pairs of atoms in opposite layers that are
stacked on top of each other, such as those appearing in AA stacking,
the orbital interaction is purely of $pp\sigma$ character. However, when atoms
are misaligned such as in SP or AB stacking configurations, the interaction is a
mixture of $pp\sigma$ and $pp\pi$ types. Tight-binding Hamiltonians have been diagonalized using the massively parallel linear algebra library ELPA  \cite{marek_elpa_2014} that allowed us to treat matrices
of order up to $N=236884$. Details on Hamiltonian matrix elements and observable calculations are discussed in \textit{Supplementary Information.}

\section{Acknowledgments}

We thank Bastien F. Grosso and Gabriel Aut\`{e}s for collaboration, Riccardo
Sabatini, Marco Gibertini and Tommaso Grioni for useful discussions.
Anton Kozhevnikov and Luca Marsella assisted us with compiling ELPA
libraries. This research was supported by the Swiss NSF grant
No. PP00P2\_133552 and Graphene Flagship. This work took advantage of the computational
facilities of the Swiss National Computing Centre (project s675).



\setcounter{equation}{0}
\setcounter{figure}{0}
\setcounter{table}{0}
\setcounter{section}{0}
\makeatletter

\renewcommand{\theequation}{S\arabic{equation}}
\renewcommand{\thefigure}{S\arabic{figure}}
\renewcommand{\thetable}{S\arabic{table}}
\renewcommand{\thesection}{S\arabic{section}}

\clearpage

\widetext

\begin{center}
\textbf{\large SUPPLEMENTARY INFORMATION}
\end{center}

\section{DFT study of interlayer interaction in bilayer graphene}

All our DFT calculations are based on the non-local rVV10 functional
implemented in QUANTUM ESPRESSO  \cite{giannozzi_quantum_2009}. This functional is composed of the revised-PW86 exchange-correlation functional plus a non-local contribution to account for the van der Waals interaction \cite{perdew_accurate_1986,murray_investigation_2009,vydrov_nonlocal_2010,sabatini_nonlocal_2013}. Ion-electron interactions have been treated by means of an ultrasoft pseudopotential
generated with the revised-PW86 functional \cite{murray_investigation_2009}. In fact, non-local contributions are not expected
to alter the effective potential generated by nuclei
and core electrons. As already found for graphite by the authors of the
rVV10 functional, relatively high values of wavefunction and charge-density
cutoffs, respectively, $E_{\mathrm{wf}}$ and $E_{\mathrm{\rho}}$, are required to describe accurately $sp^2$-carbon systems. We have used $E_{\mathrm{wf}}=80\,\mathrm{Ry}$ and $E_{\mathrm{\rho}}=574\,\mathrm{Ry}$ together with a $16\times16\times1$ Monkhorst-Pack k-point grid to
obtain converged structural observables. Periodic replicas of the system
are separated by $24~\angstrom$ of vacuum to guarantee negligible
interaction. The interlayer binding energy $E_{b}$ has been calculated
from the total energy, $E_{\mathrm{tot}}$, the energy of one isolated graphene 
layer, $E_{\mathrm{mono}}$, and the number of atoms in the unit cell,
$N_{\mathrm{c}}$, as follows
\begin{equation}
E_{b}=-\frac{1}{N_{\mathrm{c}}}\left(E_{\mathrm{tot}}-2E_{\mathrm{mono}}\right).
\end{equation}
As reported in Ref.~\cite{sabatini_nonlocal_2013} and confirmed
by our calculations, the atomic bond length of graphite is $1.42\,\mathrm{\angstrom}$
and the interlayer distance is $\Delta z_{\mathrm{graph}}=3.36\,\mathrm{\angstrom}$
in accordance with established values \cite{bacon_interlayer_1951}. Table \ref{tab:DFT} reports
the interlayer distance and the binding energy for AA and AB bilayer graphene calculated
at fixed bond length $d_{\mathrm{CC}}=1.42\,\mathrm{\angstrom}$.

\begin{table}[!h]
\centering{}\protect\caption{\label{tab:DFT}Structural observables calculated within DFT+vdW
for graphite and bilayer graphene.}
\centering{}\begin{tabular}{|c|c|c|c|}
\hline 
 & \begin{tabular}{@{}c@{}} Graphite \\  \cite{sabatini_nonlocal_2013} \end{tabular}  & Bilayer graphene - AB & Bilayer graphene - AA\tabularnewline
\hline 
\hline 
$\Delta z\left(\mathrm{\angstrom}\right)$ & $3.36$ & $3.41$ & $3.59$\tabularnewline
\hline 
$E_{b}\left(\mathrm{meV/atom}\right)$ & $39$ & $30.2$ & $25.7$\tabularnewline
\hline 
\end{tabular}
\end{table}

We have found for AB bilayer graphene the interlayer
distance $\Delta z_{\mathrm{AB}}=3.412\,\mathrm{\angstrom}$, about
$1.5\%$ larger than for graphite, consistently with previous reports \cite{mostaani_quantum_2015}. For AA stacking configuration the equilibrium distance $\Delta z_{\mathrm{AA}}=3.588\,\mathrm{\angstrom}$.
The dependencies of $E_{b}$ on $\Delta z$ and on the interlayer shift,
$\Delta x$ (see Fig.~1(b) of the main text for definition), are shown in Fig.~\ref{fig:Fit}(a,b).

\section{Determination of the classical carbon-carbon potential}

The classical pair potential for carbon atoms that we have employed
in structural relaxations consists in the sum of a short-range contribution
$V_{\mathrm{SR}}$ and a long-range contribution $V_{\mathrm{LR}}$
describing, respectively, covalent bonds and van der Waals interactions
between $sp^2$-hybridized carbon atoms. $V_{\mathrm{SR}}$ is the short-range
term of the LCBOP potential defined in Ref. \cite{los_intrinsic_2003}, adopted with no modifications.

The long-range term $V_{\mathrm{LR}}$ is a reparametrized version
of the registry-dependent potential proposed in Ref. \cite{kolmogorov_registry-dependent_2005}.
For a pair of atoms at positions $\mathbf{r}_{i}$ and $\mathbf{r}_{j}$,
with $\mathbf{n}_{i}$ ($\mathbf{n}_{j}$) being the normal vector
to the $sp^2$ hybridization plane at position $\mathbf{r}_{i}$ ($\mathbf{r}_{j}$),
$V_{\mathrm{LR}}$ is defined as

\begin{gather}
V_{\mathrm{LR}}\left(\mathbf{r}_{ij},\mathbf{n}_{i},\mathbf{n}_{j}\right)=e^{-\lambda\left(r_{ij}-z_{0}\right)}\left(C+f\left(\rho_{ij}\right)+f\left(\rho_{ji}\right)\right)-A\left(\frac{r_{ij}}{_{z_{0}}}\right)^{-6},\nonumber \\
\rho_{ij}=\left(r_{ij}^{2}-(\mathbf{n}_{i}\cdot\mathbf{r}_{ij})\right)^{1/2},\qquad\rho_{ji}=\left(r_{ji}^{2}-(\mathbf{n}_{j}\cdot\mathbf{r}_{ji})\right)^{1/2},\nonumber \\
f\left(\rho\right)=e^{-\left(\rho/\delta\right)^{2}}\sum C_{2n}(\rho/\delta)^{2n}\qquad n=0,1,2,\label{eq:Kolmogorov-Crespi}\\
\mathbf{r}_{ij}=\mathbf{r}_{i}-\mathbf{r}_{j}. \nonumber
\end{gather}

\begin{figure}[!th]
\centering{}\includegraphics[width=17cm]{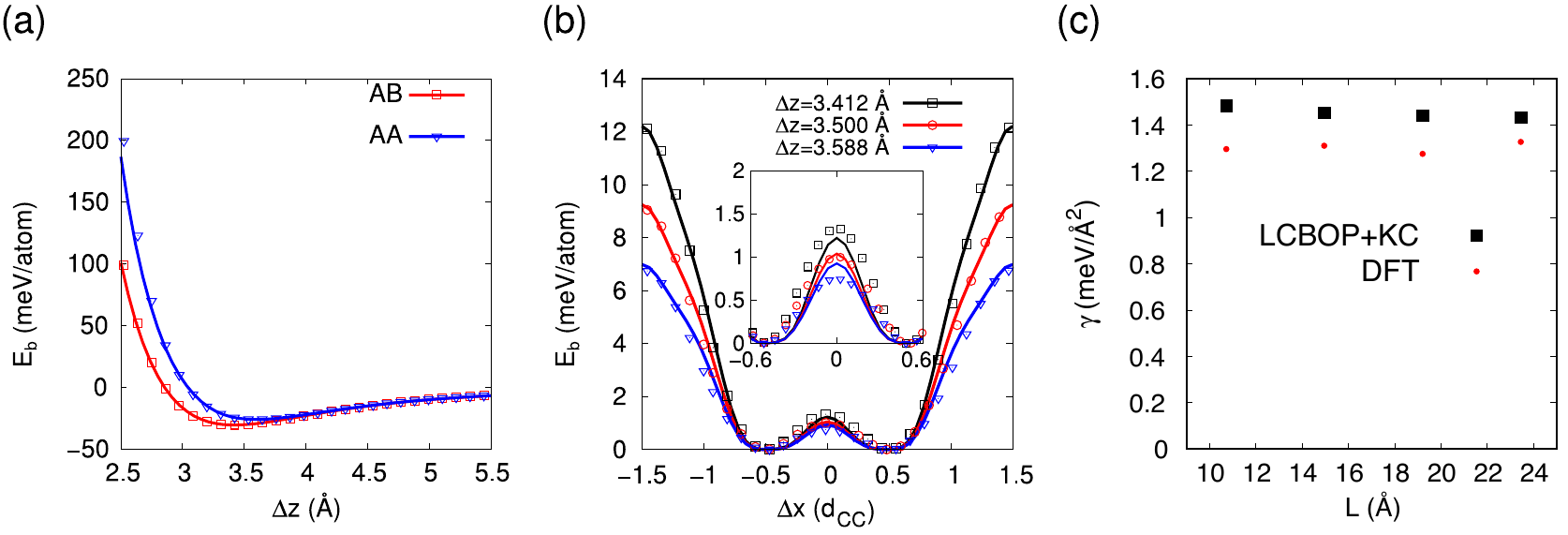}\protect\caption{\label{fig:Fit}  \textbf{a} Interlayer binding energy $E_{b}$ as a function
of interlayer distance $\Delta z$ for AA and AB-stacked bilayer graphene.
 \textbf{b} Binding energy as a function of interlayer shift $\Delta x$ (see Fig.~1(b) of the main text for definition). The energy zero is set at
$E_{b}\left(\Delta x=\pm0.5~d_{\mathrm{CC}}\right)$ corresponding
to AB/BA stacking configuration. Symbols correspond to DFT+vdW values, whereas continuous lines correspond to classical potential results. \textbf{c} Density of twist energy  $\gamma$
as a function of moir\'e periodicity $L$ calculated with DFT and LCBOP/KC.}
\end{figure}

We have made the approximation that normal vectors $\mathbf{n}_i$ are directed along the $z$ axis. Consequently,
$\rho_{ij}=\rho_{ji}=\left(\left(r_{ij}^{x}\right)^{2}+\left(r_{ij}^{y}\right)^{2}\right)^{1/2}$.
This assumption is justified by the inspection of the corrugation
of the equilibrium structures. For all relaxed models we have found that the normal vectors form an angle $\alpha<0.2^{\circ}$
with the $z$ axis. Although the original paper of Kolmogorov and
Crespi provides a set of parameters for the  potential, we have reparametrized
it by fitting Eq.~(\ref{eq:Kolmogorov-Crespi}) to three
data sets calculated within DFT+vdW. The first two datasets are the
binding energy $E_{b}$ as a function of $\Delta z$
for AB- and AA-stacked bilayer graphene (see dot data series in Fig.~\ref{fig:Fit})
and the third dataset is $E_{b}$ as a function of $\Delta x$ at fixed
interlayer distance $\Delta z=\left(\Delta z_{\mathrm{AA}}^{\mathrm{DFT}}+\Delta z_{\mathrm{AB}}^{\mathrm{DFT}}\right)/2=3.50\,\angstrom$ (red symbols in Fig.~\ref{fig:Fit}(b)). The fit has been performed
employing the non-linear optimizer DAKOTA \cite{adams_dakota_2009}.
Table~\ref{tab:Parameters-for-Kolmogorov-Crespi} compares the parameters
reported in the original reference and those resulting from our
fit.

\begin{table}[!h]
\protect\caption{\label{tab:Parameters-for-Kolmogorov-Crespi}Parameters for the Kolmogorov-Crespi
potential 
}
\hspace*{-0.85cm}
\begin{tabular}{|c|c|c|c|c|c|c|c|c|}
\hline 
 & \begin{tabular}{@{}c@{}} $C$       \\  $\left(\mathrm{meV}\right)$ \end{tabular}
 & \begin{tabular}{@{}c@{}} $C_{0}$   \\  $\left(\mathrm{meV}\right)$ \end{tabular}
 & \begin{tabular}{@{}c@{}} $C_{2}$   \\  $\left(\mathrm{meV}\right)$ \end{tabular}
 & \begin{tabular}{@{}c@{}} $C_{4}$   \\  $\left(\mathrm{meV}\right)$ \end{tabular}
 & \begin{tabular}{@{}c@{}} $z_{0}$   \\  $\left(\mathrm{\angstrom}\right)$ \end{tabular}
 & \begin{tabular}{@{}c@{}} $\delta$  \\  $\left(\mathrm{\angstrom}\right)$ \end{tabular}
 & \begin{tabular}{@{}c@{}} $\lambda$ \\  $\left(\mathrm{\angstrom}^{-1}\right)$ \end{tabular}
 & \begin{tabular}{@{}c@{}} $A$       \\  $\left(\mathrm{\angstrom}\right)$ \end{tabular}
\tabularnewline
\hline 
\hline
\begin{tabular}{@{}c@{}} x. \\  x \end{tabular}  & $3.030$ & $15.71$ & $12.29$ & $4.933$ & $3.34$ & $0.578$ & $0.578$ & $10.238$\tabularnewline

\hline 
This work & $7.183$ & $9.806$ & $5.365$ & $4.266$ & $3.516$ & $0.590$ & $3.039$ & $13.17$\tabularnewline
\hline 
\end{tabular}
\hspace*{-1cm}
\end{table}

Fig.~\ref{fig:Fit} shows a remarkable accordance between observables
calculated within DFT (symbols data series) and KC potential (continuous
curves). In particular, the equilibrium distances calculated by means
of the reparametrized KC potential are $\Delta z_{\mathrm{AA}}^{\mathrm{KC}}=3.599\,\mathrm{\angstrom}$
and $\Delta z_{\mathrm{AB}}^{\mathrm{KC}}=3.416\,\mathrm{\angstrom}$,
and the respective binding energies for AA and SP stacking configurations calculated at $\Delta z=3.412\,\mathrm{\angstrom}$ are $E_{b}^{\mathrm{AA}}=12.2\,\mathrm{meV/atom}$ and $E_{b}^{\mathrm{SP}}=1.22\,\mathrm{meV/atom}$ relative to AB stacking configuration. Finally, the in-plane carbon-carbon
bond length is $d_{\mathrm{CC}}=1.419\,\angstrom$.

As a test of the transferability of the classical potential resulting from our fit, in Fig.~\ref{fig:Fit}(c) we show a comparison of the density of twist energy calculated within DFT and LCBOP/KC for several bilayer graphene models. In particular, we focus on systems with small moir\'e periodicity $10~\angstrom<L<24~\angstrom$ (i.e. large twist angles $6^\circ<\theta<13^\circ$) in their equilibrium configuration obtained by atomic relaxation. Treating the interlayer interaction classically introduces a discrepancy that is smaller than 10\% in most of the cases. 

\section{Structural relaxation}

\begin{figure}[!th]
\centering{}\includegraphics[width=14cm]{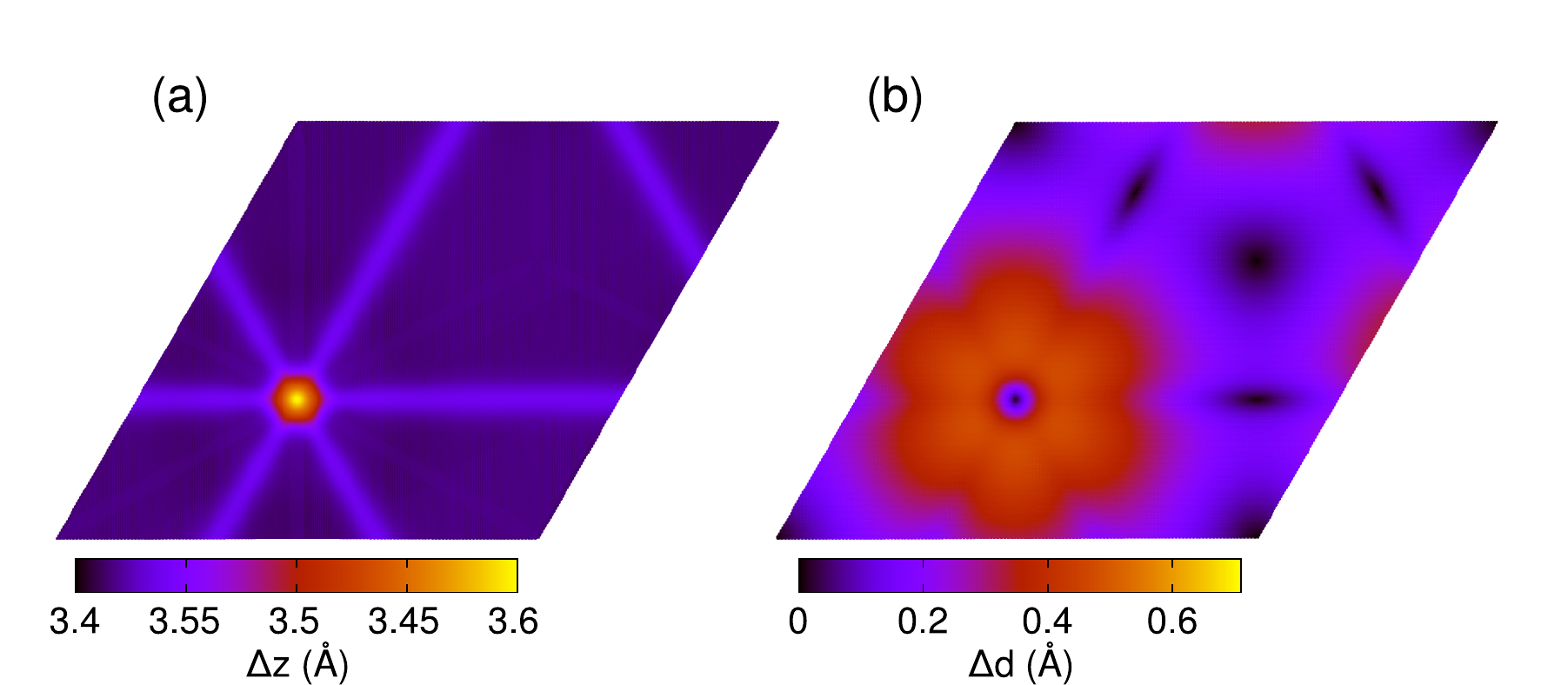}\protect\caption{\label{fig:DIFF} \textbf{a} Maps of the interlayer distance $\Delta z$ and \textbf{b} the absolute magnitude of the atomic displacement $\Delta d$ upon relaxation of a
rigidly twisted BLG model characterized by the twist angle $\theta=0.235^{\circ}$, corresponding to the moir\'{e} periodicity $L=59.8~\mathrm{nm}$.
The upper bound for the atomic displacement is $d_{\mathrm{CC}}/2=0.71\,\mathrm{\angstrom}$,
as explained in the main text.}
\end{figure}

The full potential $V=V_{\mathrm{SR}}+V_{\mathrm{LR}}$ has been implemented
in LAMMPS  \cite{plimpton_fast_1995,lammps_2015}. Twisted BLG structures
have been initially relaxed by means of conjugate gradient plus
quadratic line search method \cite{fletcher_practical_2013} and fine minimization was obtained using fast inertia relaxation method \cite{bitzek_structural_2006}. The supercell
vectors have been kept fixed and the initial interlayer distance has been set to $\Delta z^{\mathrm{KC}}_{\mathrm{AB}}=3.416\,\mathrm{\angstrom}$. At the end of the relaxation the largest
force component acting on any atom was below $3\,\mathrm{meV}/\mathrm{\angstrom}$.
In Fig.~\ref{fig:DIFF}, the full maps of the interlayer distance
and in-plane atomic displacement are shown. With respect to Fig.~3(c,d) of the main text,
the full maps allow to appreciate the whole set of symmetries. $\Delta z$
is almost constantly equal to $\Delta z_{\mathrm{SP}}^{\mathrm{KC}}=3.439\,\angstrom$
along the soliton lines and to $\Delta z_{\mathrm{AB}}^{\mathrm{KC}}=3.416\,\angstrom$
in AB/BA domains.

\section{Electronic structure calculations}

In this section, we present the details of our electronic structure calculations.
The tight-binding model Hamiltonian for bilayer graphene has been taken
from Ref.~\cite{trambly_de_laissardiere_localization_2010} (see
also Ref.~\cite{trambly_de_laissardiere_numerical_2012} for thorough
electronic structure calculations of rigidly twisted bilayer graphene).
In such a model, only $p_{z}$-orbitals for carbon atoms are considered. The Hamiltonian is defined as
\begin{equation}
H=\sum_{i\neq j}V_{ij}a_{i}^{\dagger}a_{j}\label{eq:Hamiltonian},
\end{equation}
where the operators $a_{i}^{\dagger}$ and $a_{i}$, respectively, create and annihilate
an electron in the $p_{z}$-orbital of the atom at position $\mathbf{R}_{i}$.
The matrix elements, $V_{ij}$, are obtained by combining $\sigma$-
and $\pi$-type Slater-Koster parameters $V_{pp\sigma}$ and $V_{pp\pi}$
in the approximation that the axes of $p_{z}$-orbitals are parallel, akin
to the assumption that the normal vectors of the two graphene layers
are parallel. One has 
\begin{equation}
V_{ij}=V_{pp\pi}\sin^{2}\left(\theta\right)+V_{pp\sigma}\cos^{2}\left(\theta\right),\label{eq:SK}
\end{equation}
where $\theta$ is the angle between the orbital axes and the vector
$\mathbf{R}_{ij}=\mathbf{R}_{i}-\mathbf{R}_{j}$ connects the
two orbital centers \cite{slater_simplified_1954}. For a pair of atoms in the same layer $\theta=90^{\circ}$
and $V_{ij}=V_{pp\pi}$. Conversely, for a pair of atoms placed on top of each other in opposite layers, namely forming a dimer, $\theta=0^{\circ}$
and $V_{ij}=V_{pp\sigma}$. $V_{pp\pi}$ and $V_{pp\sigma}$
depend  exponentially on the distance between the two orbital centers $r=\left|{\bf R}_i-{\bf R}_j\right|$ as
\begin{equation}
V_{pp\pi}\left(r\right)=V_{pp\pi}^{0}e^{q_{\pi}(1-r/a_{\pi})},\qquad V_{pp\sigma}\left(r\right)=V_{pp\sigma}^{0}e^{q_{\sigma}(1-r/a_{\sigma})}.\label{eq:matrix_elements}
\end{equation}

Following Ref.~\cite{trambly_de_laissardiere_localization_2010}
we assume $V_{pp\pi}^{0}=-2.7\,\mathrm{eV}$, $V_{pp\sigma}^{0}=0.48\,\mathrm{eV}$,
$a_{\pi}=1.419\,\mathrm{\angstrom}$, $q_{\pi}=3.1454$. Differently from Ref.~\cite{trambly_de_laissardiere_localization_2010}, in order to be consistent with the interlayer distance for AB stacking configuration
calculated in the present work, we have taken $a_{\sigma}=3.417\,\mathrm{\angstrom}$
and $q_{\sigma}=8.200$. The long distance cut-offs for $V_{pp\sigma}\left(r\right)$
and $V_{pp\pi}\left(r\right)$ have been fixed, respectively, at $\bar{r}_{\sigma}=3.5\,\mathrm{\angstrom}$
and $\bar{r}_{\pi}=5\,\mathrm{\angstrom}$. We have verified that
further increasing these cut-offs does not affect the calculated
observables. In all our calculations the charge neutrality point, corresponding to the Fermi energy for undoped systems, is $E_{f}=0.82\,\mathrm{eV}$.

Density of states are calculated as follows
\begin{equation}
\mathrm{DOS}\left(E\right)=\int_{\mathrm{BZ}}\mathrm{d}\mathbf{k}\sum_{n_{\mathbf{k}}}\delta\left(E-E_{n_{\mathbf{k}}}\right),\label{eq:DOS}
\end{equation}
with $n_{\mathbf{k}}$ running over all the eigenvalues at position
$\mathbf{k}$ in reciprocal space. For computational needs the
$\delta$-function appearing in eq.~\ref{eq:DOS} has been replaced by
a Lorentzian function:
\begin{equation}
\delta\left(E-E_{n_{\mathbf{k}}}\right)\rightarrow\frac{1}{\pi}\frac{\eta}{\left(E-E_{n_{\mathbf{k}}}\right)^{2}+\eta^{2}},\qquad\eta\rightarrow0^{+}.\label{eq:lorentzian}
\end{equation}

The local density of states on the $i$-th atom at position $\mathbf{R}_{i}$,
integrated in the energy range $\left[E_{1},E_{2}\right]$, has been calculated
as follows
\begin{equation}
\mathrm{LDOS}\left(i;E_{1},E_{2}\right)=\int_{E_{1}}^{E_{2}}\mathrm{d}E=\int_{\mathrm{BZ}}\mathrm{d}\mathbf{k}\sum_{n_{\mathbf{k}}}\delta\left(E-E_{n_{\mathbf{k}}}\right)\left|\left\langle i\left|n_{\mathbf{k}}\right.\right\rangle \right|^{2}=\int_{\mathrm{BZ}}\mathrm{d}\mathbf{k}\sum_{E_{1}<E_{n_{\mathbf{k}}}<E_{2}}\left|\left\langle i\left|n_{\mathbf{k}}\right.\right\rangle \right|^{2},\label{eq:LDOS}
\end{equation}
where $\left | i \right>$ represents the $p_z$ orbital of the $i$-th atom. Spin has not been explicitly considered in our calculations.

Integration over the Brillouin zone have been performed introducing a discrete
grid. The band structures shown in panels (a) and (b) of Fig.~5 of the main text have been calculated with a Monkhorst-Pack grid of $25\times25$ and $5\times5$ k-points, respectively. The density of states of the model shown in Fig.~6 of the main text has been calculated using the high-symmetry points $\Gamma$ and $\mathrm{M}$ of
the hexagonal Brillouin zone, taking advantage of the fact that
the Hamiltonian represented in reciprocal space $H\left(\mathbf{k}=\mathrm{M},\Gamma\right)$
is a real matrix.

Matrix diagonalizations have been performed employing the Eigenvalue Solvers
for Petaflop Applications library (ELPA) \cite{marek_elpa_2014}. This allowed
us to diagonalize a $N\times N$ real matrix with $N=236884$ in about $2.5$
hours using $1024$ CPU cores.

\FloatBarrier


%

\end{document}